\documentclass[twocolumn,a4paper,nofootinbib]{revtex4} 
\usepackage{graphicx}
\usepackage[breaklinks=true,colorlinks=true,linkcolor=blue,urlcolor=blue,citecolor=blue]{hyperref}

\usepackage{color}

\usepackage[utf8x]{inputenc}
\usepackage[nointegrals]{wasysym}
\usepackage{tikz}
\usepackage{array}
\usepackage{verbatim}
\usepackage{amsmath}
\usepackage{xspace}
\usepackage{wrapfig}

\usepackage{mathrsfs}
\DeclareSymbolFontAlphabet{\mathrsfs}{rsfs}
\newcommand{\scri}{\mathrsfs{I}}
\newcommand{\scripx}{$\scri^+$}
\newcommand{\scrip}{\scripx\xspace}

\newcommand{\papfirst}{Vano-Vinuales:2014koa}
\newcommand{\papgauge}{Vano-Vinuales:2017qij}

\newcommand{\alexthesis}{Vano-Vinuales:2015lhj}
\newcommand{\procere}{Vano-Vinuales:2014ada}
\newcommand{\procmg}{Vano-Vinuales:2016mbo}

\newcommand{\eref}[1]{(\ref{#1})}
\newcommand{\sref}[1]{section~\ref{#1}}
\newcommand{\ssref}[1]{subsection~\ref{#1}}
\newcommand{\fref}[1]{figure~\ref{#1}}

\newcommand{\Kc}{K_{\textrm{CMC}}}

\newcommand{\atscrip}[1]{\left.#1\right|_{\scri^+}}

\newcommand{\fl}{\hspace*{-\mathindent}}

\begin{document}

\title[Free hyperboloidal evolution in 3D]{Free hyperboloidal evolution in 3D: gauge waves and constraint-violating gravitational waves} 

\author{Alex Vañó-Viñuales}
\address{Departament de Física, Universitat de les Illes Balears, IAC3, Carretera Valldemossa km 7.5, E-07122 Palma, Spain}

\begin{abstract}

Hyperboloidal evolution allows to reach future null infinity within numerical relativity simulations. This work extends the conformal compactification approach with BSSN/Z4 from spherical symmetry to full 3D simulations. 
This is the first free numerical hyperboloidal evolution in 3D with a common formulation of general relativity. 
In this first step, gauge waves and constraint-violating gravitational waves perturbing Minkowski spacetime are evolved nonlinearly and show that the implementation is reliable and signals can be successfully extracted at future null infinity.

\end{abstract}

\maketitle


\section{Introduction}

\begin{figure}[h!!!!] 
    \centering
    \includegraphics[width=0.48\linewidth]{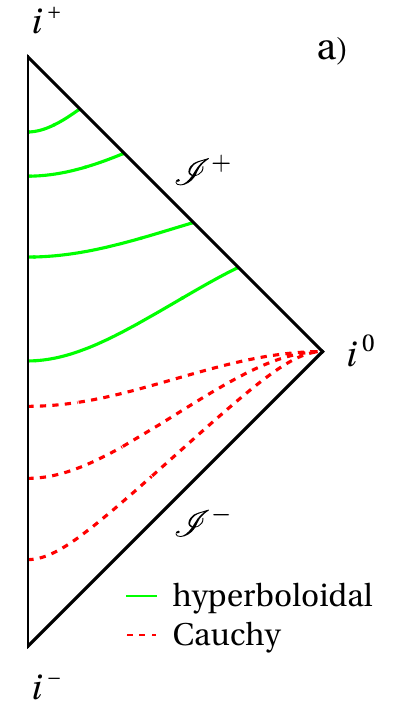}
    \includegraphics[width=0.48\linewidth]{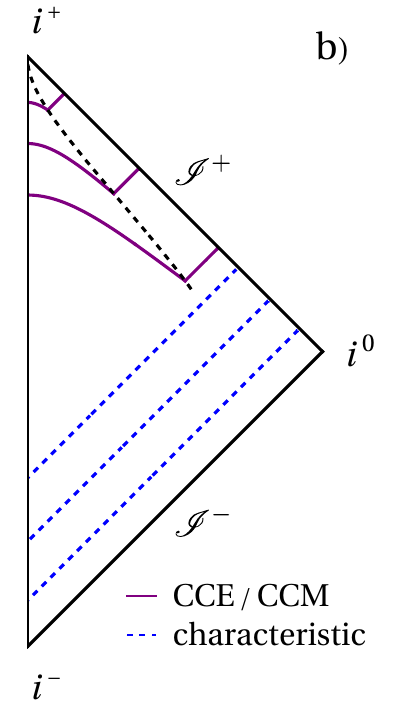}
    \caption{Slices of Minkowski spacetime: a) hyperboloidal and Cauchy, and b) Cauchy-characteristic and characteristic.}\label{pendiags}
\end{figure}

Within the modelling of isolated systems, such as binaries emitting gravitational waves (GWs), the far-field zone encodes crucial information of the spacetime. Future null infinity  (\scrip, see \fref{pendiags}) corresponds to the collection of the end points of future-directed null geodesics and thus plays a fundamental role, from both a mathematical and numerical perspective. On one hand, global quantities like the total energy of a spacetime and the emitted gravitational radiation are only unambiguously defined there. On the other hand, \scrip is also the idealized location of astrophysical observers \cite{Barack:1998bv,Leaver1986,PhysRevD.34.384}, such as GW interferometers on Earth or in space. Thus, it is at future null infinity that GW signals from numerical simulations of compact binary coalescences are to be extracted, to provide waveforms as accurate as possible for the analysis of GW interferometry data. 

There are several possibilities to extract GWs from numerical relativity simulations. 
The simplest is to extract Cauchy-evolved data at several finite values of the radial coordinate and extrapolate \cite{Boyle:2009vi,Bishop:2016lgv,Iozzo:2020jcu} them to \scrip. Although widely used, this method is neither accurate nor suitable enough, as it contains systematic errors \cite{Hinder:2013oqa,Taylor:2013zia} and does not provide the GW memory effect \cite{Mitman:2020pbt} in the waveforms. 

Characteristic evolution, where the data are given along null slices \cite{PhysRevLett.80.3915,Winicour2009LRR} (see the blue dashed lines in \fref{pendiags}b), allows to reach future null infinity straightforwardly. This simple approach has become very popular in spherical symmetry, see e.g.~\cite{1983RSPSA.386..373C,1999.Christodoulou,PhysRevD.36.3575,Gomez199211,PhysRevD.51.5558,Husa:2000kr}, and it is also a common choice within studies of criticality at the threshold of gravitational collapse such as \cite{Santos:2025epw,Madler:2026fve}. Still, the curvature from the strong field can create caustics, a breakdown of the coordinate system that needs to be treated \cite{stefried,1983RSPSA.385..345F}. 

Evolving on usual Cauchy spacelike slices (red dashed lines in \fref{pendiags}a) up to a given radius and joining on to characteristic slices raises two options: Cauchy-characteristic extraction/evolution (CCE) and Cauchy-characteristic matching (CCM), whose setup is shown in purple solid lines on \fref{pendiags}b. 
In CCE \cite{Bishop:1996gt,Reisswig:2009us,Babiuc:2011qi,Moxon:2021gbv,Ma:2024bed,DaRe:2026wyd} the Cauchy evolution is performed independently and used as inner boundary data for the characteristic one. This is currently the most widely-used state-of-the-art approach for obtaining signals at \scrip, but simulations are limited in time by the finite-radius outer boundaries of the Cauchy evolution. 
CCM \cite{BishopCCM,Szilagyi:2000xu,Ma:2023qjn,Ma:2024hzq}, in which the Cauchy and characteristic domains are evolved in parallel, does not have a time limitation. However, a critical aspect is well-posedness of the complete construction: the Bondi-like formulation used for the characteristic part is ill-posed (only weakly hyperbolic) beyond spherical symmetry \cite{Giannakopoulos:2020dih,Giannakopoulos:2021pnh}, meaning that there may be configurations for which CCM will not converge. As alternatives there are tetrad-based formulations \cite{Ripley:2021wyy} or systems where the curvature is also evolved \cite{Racz:2013hva,Cabet:2014sha}. 

A more flexible and elegant alternative is hyperboloidal evolution \cite{friedrich1983,Friedrich86}, which uses spacelike slices that reach null infinity (see green solid lines in \fref{pendiags}a). Their main advantages are their smoothness and the absence of matching interfaces or abrupt changes in their causality structure and formulations used, as well as their suitability to be employed with a 3+1 decomposition of the Einstein equations (solving as initial value problem). This approach is taken here. 

A suitable method for hyperboloidal evolution is conformal compactification, based on Penrose's idea \cite{PhysRevLett.10.66} and given by~\eref{conftrafog}. Here, null infinity is reached with a finite coordinate value, at the cost of formally singular equations there. The mathematical foundation on the conformally compactified hyperboloidal initial value problem \cite{Frauendiener:1997ze} was laid with the development of the Conformal Field Equations \cite{Fredrich:aaa}, shown to be manifestly regular \cite{friedrich1983,Friedrich:2003fq}. 
Hyperboloidal foliations are also used for proving global existence, e.g.~\cite{doi:10.1142/9427,LeFloch:2015wqa,LeFloch:2017yjt,LeFloch:2023xij}.
 
The first numerical implementations of the Conformal Field Equations in a metric-based approach were \cite{Hubner:1994pd,Hubner:1998hn,Hubner:1999th,Hubner:2000zn,Hubner:2000pb}, with continuum instabilities found later\cite{Husa:2002kk,Husa:2005ns,Husa:2002zc,Frauendiener:2004bj}. 
A tetrad formalism was pursued \cite{Frauendiener:1997zc,Frauendiener:1997ze,Frauendiener:1998yi} within axisymmetry. 
This approach currently uses the COFFEE code \cite{Doulis:2019ejj}, and has produced relevant work in full general relativity \cite{Frauendiener:2021eyv,Frauendiener:2022bkj,Frauendiener:2023ltp,Camden:2025xjv,Frauendiener:2025xcj}. The disadvantage of this approach for gravitational wave extraction is that the gauge used (conformal Gauss gauge) compactifies the time coordinate and features cannot be resolved at late times, and so far it is only vacuum-compatible. Also, it does not allow for ``scri-fixing'' -- fixing \scrip to a specific location in the spatial grid, a very desirable feature for hyperboloidal evolution. Steps towards achieving this with the Conformal Field Equations have been taken with the linearized theory in \cite{Peterson:2025kfb}. 

A milestone in the numerical work was put by Zengino\u{g}lu, with Schwarzschild black hole evolutions using conformally compactified generalized harmonic gauge in spherical symmetry \cite{Zenginoglu:2007it}. This work set the basic ingredients for free evolution on hyperboloidal slices with a time-independent conformal factor \cite{Zenginoglu:2007jw,Zenginoglu:2008pw,Zenginoglu:2008wc}.
A formalism \cite{Moncrief:2008ie,Rinne:2013qc} making use of the computationally slower constrained evolution resulted in an axisymmetric code \cite{Rinne:2009qx}. A tetrad formalism \cite{Bardeen:2011ip} has been used in the also constrained evolutions \cite{Morales:2016rgt}.  

An approach other than conformal compactification that also evolves hyperboloidal slices reaching \scrip is the dual frame formalism \cite{Hilditch:2015qea,Hilditch:2016xzh}. Its suitability to treat the far-field region has been shown with extensive mathematical analyses \cite{Gasperin:2018tvk,Duarte:2021xtz,Duarte:2022mxj,Duarte:2022vxn,Duarte:2022jpg,Duarte:2025nxc}, numerical work with toy models \cite{Gasperin:2019rjg,Peterson:2023bha}, including the development of Summation-By-Parts algorithms \cite{Gautam:2021ilg,Gautam:2026cch,Reddy:2026dfz}, and spherically symmetric numerical evolutions of the Einstein equations \cite{Peterson:2024bxk,Peterson:2025csg}.
%
Yet another recently developed method follows the formulation in \cite{Nutzi:2025kqc}, has been applied to semilinear waves in  \cite{Zenginoglu:2026bwc} and is pushing for Cauchy-hyperboloidal extraction, with related prior work \cite{Bernuzzi:2025hhu}. 

Hyperboloidal initial data including black holes has been studied in \cite{Schneemann,Buchman:2009ew,Bardeen:2011pd,Schinkel:2013zm,Schinkel:2013tka,schinkelthesis}, with spherically symmetric work on data with a neutron star in \cite{Vano-Vinuales:2025oai}, but more work is still needed for the desired compact binary configurations, for instance in formulations \cite{Beyer:2017qjo,Csukas:2025eer}. 

Besides the full non-linear case, hyperboloidal slices as background have become the standard method for the Teukolsky equation \cite{Zenginoglu:2011zz,Harms:2014dqa}, extended to the conformal Teukolsky equation \cite{Gasperin:2026znw}, and used with pseudospectral methods \cite{Zenginoglu:2010zm,Ansorg:2016ztf}. They are common in quasi-normal mode studies and the pseudospectrum \cite{PanossoMacedo:2018hab,Jaramillo:2020tuu,Destounis:2021lum,Bourg:2024jme,PanossoMacedo:2024nkw}, and of special interest for numerical scattering \cite{Baptista:2025kmo,Demirboga:2026fpa,Barrer:2026xgt}. 


The present approach uses conformal compactification and a time-independent conformal factor together with free evolution of BSSN~\cite{PhysRevD.52.5428,Baumgarte:1998te} and Z4~\cite{bona-2003-67} formulations, which are known to  successfully evolve black hole spacetimes with singularity-avoiding slicing conditions \cite{bona}. 
Previous work in spherical symmetry includes the main formulation \cite{\papfirst,\alexthesis} with a study on gauge conditions \cite{\papgauge}, black hole initial data and evolutions \cite{\procere,\procmg,Vano-Vinuales:2023yzs,Vano-Vinuales:2023pum}, evolution of different hyperboloidal slicings \cite{Vano-Vinuales:2024tat} that include the first hyperboloidal layer \cite{Zenginoglu:2010cq} evolutions, and the addition of electromagnetism and charged scalar fields \cite{Alvares:2025pbi,Alvares:2025jjm}. 

Here I present the first step beyond spherical symmetry with this approach. To the author's best knowledge, these are the first free 3D metric-based hyperboloidal evolutions of general relativity with a commonly used formulation and complete hyperboloidal slices. 
Including future null infinity in numerical simulations is motivated by the need to significantly improve how compact-object binary simulations are performed: what we want are efficient codes that can robustly run for as long as needed and provide reliable and accurate GW signals that will be suited for the next generation of GW interferometers. 



This paper is structured as follows: \sref{setup} covers the formulation and gauge conditions used, as well as the initial data considered and the implementation in the code. The results are presented in \ref{results}, and in \sref{conclusions} they are put in perspective and the next steps are discussed. The metric's signature is $(-,+,+,+)$, and natural units $G=c=1$ are used.

\section{Setup}\label{setup}

\subsection{Formulation}\label{formulation}
Conformal compactification uses a conformally rescaled metric $\bar g_{ab}$ instead of the physical one $\tilde g_{ab}$, because the latter diverges at null infinity after a compactification of the (spatial) coordinates on a hyperboloidal slice. The relation between metrics is given by
\begin{equation}\label{conftrafog}
\bar g_{ab} =\Omega^2\tilde g_{ab}, 
\end{equation}
where the conformal factor $\Omega$ is taken to be time-independent and satisfies that $\Omega|_{\scri^+}=0$ to keep $\bar g_{ab}$ finite at null infinity, with $\bar\nabla_a\Omega|_{\scri^+}\neq0$ also holding. 

The Einstein equations for the physical metric $\tilde g_{ab}$, including constraint propagation and damping terms of the Z4 formalism \cite{bona-2003-67,Bona:2003qn,Gundlach:2005eh}, are 
\begin{align}\label{einsteinphyseq}
G[\tilde g]_{ab}&+2\tilde\nabla_{(a} Z_{b)} -\tilde g_{ab}\tilde\nabla^cZ_c \nonumber\\&- \kappa_1\left(2\,\tilde n_{(a}Z_{b)}+\kappa_2\,\tilde g_{ab}\,\tilde n^cZ_c\right) =8\pi T_{ab}, 
\end{align}
where $G[\tilde g]_{ab}$ in the Einstein tensor constructed from the physical metric, $T_{ab}$ is the stress-energy-momentum tensor, $\tilde n^a$ is the timelike normal vector and $\kappa_1$ and $\kappa_2$ are given parameters. The Einstein equations are satisfied when the Z4 field $Z_a$ vanishes.
In terms of the conformally rescaled metric $\bar g_{ab}$ as defined in \eref{conftrafog} and the rescaled normal vector $\bar n^a\equiv\tilde n^a/\Omega$, the Einstein equations become 
\begin{align}\label{einsteineq}
\fl G[\bar g]_{ab}& + {2\over\Omega}(\bar\nabla_a\bar\nabla_b\Omega-\bar g_{ab}\bar\Box \Omega) + {3\over\Omega^2}\bar g_{ab}(\bar\nabla_c\Omega)(\bar\nabla^c\Omega)
\nonumber\\&+2\bar\nabla_{(a} Z_{b)} -\bar g_{ab}\bar\nabla^cZ_c + \frac{4}{\Omega} Z_{(a}\bar\nabla_{b)}\Omega \nonumber\\&- \frac{\kappa_1}{\Omega}\left(2\,\bar n_{(a}Z_{b)}+\kappa_2\,\bar g_{ab}\,\bar n^cZ_c\right)=8\pi T_{ab}  . 
\end{align}
The first line corresponds to the Einstein tensor of the physical metric, $G[\tilde g]$, expressed in terms of that of the conformal metric, $G[\bar g]$, plus formally divergent terms at \scrip. Here indices are raised and lowered with the conformal metric $\bar g_{ab}$, its covariant derivative is denoted by $\bar\nabla$ and $\bar\Box\equiv\bar g^{ab}\bar\nabla_a\bar\nabla_b$.

The above conformally rescaled Einstein equations are 3+1-decomposed and expressed in the conformal Z4 \cite{bona-2003-67,Alic:2011gg,Sanchis-Gual:2014nha} formulation in the form of \cite{Bernuzzi:2009ex}.  
The line element is given by
\begin{equation}
    \begin{aligned}
    d\bar s^2 = -\alpha^2 dt^2 + \chi^{-1}\gamma_{ij}(dx^i + \beta^i dt)(dx^j + \beta^j dt)\,,
    \end{aligned}
    \label{eq:lineelement}
\end{equation} 
where $\alpha$ is the lapse and $\beta^i$ is the shift vector, the gauge quantities that are evolved besides $\gamma_{ij}$ and $\chi$, the twice conformally rescaled spatial metric and its conformal factor respectively. The conformal extrinsic curvature\footnote{The convention for the sign of the extrinsic curvature is that of Misner, Thorne, and Wheeler \cite{Misner1973}, meaning that a negative value means expansion of the normals.} tensor $\bar{K}_{ij}$ is divided into its trace and trace-free parts,
\begin{equation}
    \bar{K} = \chi{\gamma}^{ij}\bar{K}_{ij}, \qquad A_{ij} = \chi \bar{K}_{ij} - \frac{1}{3}\gamma_{ij}\bar{K}\,,
\end{equation}
Instead of $\bar{K}$, the evolved quantity is $\Delta\tilde{K} = \tilde K -\Kc$, with $\Kc$ a given parameter of the hyperboloidal slice and $\tilde K$ the physical trace of the extrinsic curvature mixed with the physical $\tilde \Theta=\Omega\Theta$, a Z4 evolution variable: 
\begin{equation}
    \tilde{K} = \Omega \bar{K} - 2 \tilde \Theta - \frac{3 \beta^i \partial_i \Omega}{\alpha}. 
    \label{eq:ktransformations}
\end{equation} 
Finally, the Z4 quantity $Z_i$ is introduced into the evolution variable
\begin{equation}\label{Lambdadef}
    \Lambda^i = \Delta \Gamma^i + 2 \gamma^{ij}Z_j\,,
\end{equation}
with $\Delta \Gamma^i = \gamma^{jk}\left(\Gamma^i_{jk}-\hat\Gamma^i_{jk}\right)$. Here $\hat{\Gamma}_{jk}^i$ is the connection constructed from a chosen time-independent background spatial metric $\hat\gamma_{ij}$, which is simply set to the flat spatial metric. Thus, the evolution variables are $\gamma_{ij}$, $\chi$, $\alpha$, $\beta^i$, $A_{ij}$, $\Delta\tilde{K}$, $\Lambda^i$ and $\tilde{\Theta}$. 
The implemented equations correspond to (2.66) and (2.68) in \cite{\alexthesis}, and from now on we will consider a vacuum spacetime, namely $T_{ab}=0$. 

\subsection{Gauge conditions}\label{gauge}

The hyperbolic gauge conditions considered for the simulations in this work are a hyperboloidal-adapted version of the generalized harmonic gauge conditions. They make scri-fixing \cite{Frauendiener:1997ze,Zenginoglu:2007jw} possible, which allows to set \scrip at the boundary of the numerical grid and so avoid using any physical boundary conditions there. This is because \scrip is an ingoing null surface and no physical information is allowed to enter through it. A potential alternative to scri-fixing is introducing a small cosmological constant \cite{vanMeter:2006mv}, as has been applied to scalar waves in \cite{Zenginoglu:2026gqq}. 


These gauge conditions also satisfy the preferred conformal gauge~\cite{Tamburino:1966zz} at \scrip, namely $\atscrip{\bar\Box\Omega}=0$. They are described in detail in section 4 in~\cite{\papgauge} and can be derived by isolating
\begin{align}\label{pbg}
&\partial_t\alpha \textrm{ and }\partial_t\beta^i \quad\textrm{ from }\quad \tilde g^{ab}\left(\tilde\Gamma^c_{ab}-\hat{\tilde{\Gamma}}^c_{ab}\right) = \tilde F^c, 
\end{align}
with the choice of source functions 
\begin{align}\label{ffuncs}
\tilde F^t = \xi_\alpha\Omega\frac{\alpha-\hat\alpha}{\alpha^3}\quad \textrm{ and } \quad\tilde F^i = 0, 
\end{align}
where $\xi_\alpha$ is a free parameter. This form of $\tilde F^t$ is different from that used in~\cite{\papgauge}, has better convergence properties at \scrip, and this is the first work where it is used. 
Besides this important point, these harmonic-like gauge conditions propagate with the speed of light and have been tested in spherical symmetry with successful results, so they were the first choice for the full 3D tests. 
The explicit expressions for~\eref{pbg} above are included in appendix~\ref{appendix}. 

\subsection{Initial data}\label{inidata}

The initial data for the evolution variables are given on a $t=constant$ slice, where $t$ is the hyperboloidal time, and corresponds to a constant-mean-curvature slice of (perturbed) Minkowski spacetime with $\tilde K=\Kc$. 
The physical radial coordinate $\tilde r$ on the slice is compactified to reach \scrip with a finite value of the compressed $r$, \begin{equation}
    \tilde{r} = \frac{r}{\Omega},
    \label{eq:compacfactor}
\end{equation}
with compactification and conformal factor given by
\begin{equation}
    \Omega = -\Kc\frac{r^2_{\scri^+}-r^2}{6r_{\scri^+}}.
    \label{eq:confexp}
\end{equation}

The values for the time-independent background metric quantities and the initial data for all evolution variables take the following form
\begin{gather}
    \hat\alpha = \sqrt{\Omega^2 + \left( \frac{\Kc r}{3}\right)^2}, \ \alpha_0 = \hat\alpha+G, \quad \hat\chi=\chi_0 = 1, \nonumber\\
    \hat\gamma_{ij}= \left(\begin{array}{ccc}1&0&0\\0&r^2&0\\0&0&r^2\sin^2\theta\end{array}\right), \  \gamma_{ij0}= \left(\begin{array}{ccc}e^{\frac{2}{3}B}&0&0\\0&r^2e^{\frac{2}{3}B}&0\\0&0&\frac{r^2\sin^2\theta}{e^{\frac{4}{3}B}}\end{array}\right),\nonumber\\
    \hat\beta^i = \beta^i_0 =\left(\begin{array}{c}\frac{K_{\text{CMC}}r}{3}\\0\\0\end{array}\right), \ \begin{array}{cc} \Delta\tilde K_0=0,& \tilde\Theta_0 = 0,\\A_{ij0} = 0_{ij}, & \Lambda^i_{0} = \Delta \Gamma^i_0,\end{array} 
    \label{eq:indatagen3}
\end{gather}
where $0_{ij}$ denotes the 6x6 zero-matrix and $\Delta \Gamma^i_0=\gamma^{jk}_0\left(\Gamma^i_{jk0}-\hat\Gamma^i_{jk}\right)$ is calculated from the $\gamma_{ij0}$ and $\hat\gamma_{ij}$ given above. Perturbations are introduced via $G=G(r,\theta,\varphi)$ in $\alpha_0$ for constraint-satisfying gauge wave initial data, and via $B=B(r,\theta,\varphi)$ in $\gamma_{ij0}$ for Brill-like constraint-violating GW initial data. 

Brill waves \cite{BRILL1959466} describe a solution of the constraint equations in axisymmetric vacuum at a moment of time symmetry, where the extrinsic curvature vanishes identically. Here the perturbation is described as Brill-like for its structure (see e.g.~equations (5) and (6) in~\cite{Fernandez:2022hyx}) within the metric, satisfying the condition that $\gamma_{ij}$ has determinant equal to $r^4\sin^2\theta$. However a hyperboloidal slice needs a non-vanishing extrinsic curvature in the asymptotic regime, so the solution will not be time-symmetric. In the hyperboloidal slices of Minkowski considered here as initial data, the trace-free part of the extrinsic curvature vanished, but its trace is constant (and equal to $\Kc$). Besides, for simplicity the constraints are not solved for the Brill-like perturbations included. Still, it provides a non-vanishing Weyl scalar $\Psi_4$ close to \scrip, indicating that it carries GW content. 

The form of the initial perturbation is 
\begin{equation}
A e^{-\frac{(r^2-c^2)^2}{4\sigma^4}}\ F_\theta(\theta) \ F_\varphi(\varphi), 
\end{equation}
where $A$, $c$ and $\sigma$ are given parameters. 
\begin{itemize}\itemsep0ex
\item For the 3D simulations presented in \ssref{3d}, $F_\theta(\theta)=\sin^2\theta$ and $F_\varphi(\varphi)=\sin(2\varphi)$. This angular setup is proportional to the imaginary part of the $Y_2^2(\theta,\varphi)$ spherical harmonic. 
\item For the axially symmetric runs shown in \ssref{conv}, $F_\varphi(\varphi)=1$, and $F_\theta(\theta)=3\cos^2\theta-1$ for gauge waves and $F_\theta(\theta)=\sin^2\theta$ for constraint-violating GWs. The inspiration is the axisymmetric $Y_2^0(\theta,\varphi)$ spherical harmonic, which worked well for gauge perturbations. However, in the axisymmetric constraint-violating Brill-like GW simulations, convergence was lost at the extrema of the $\theta\in(0,\pi)$ range (most likely because of the use of curvilinear coordinates), so choosing a function vanishing at those extrema like $\sin^2\theta$ provided a better behaviour. 
\end{itemize}
For the slice and compactification the parameters are set to $\Kc=-1$ and $r_{\scri^+}=1$, while those in the equations are $\kappa_1=1.5$, $\kappa_2=0$ and $\xi_\alpha=0.7$. The initial perturbation uses $A=0.05$, $c=0.75$ and $\sigma=0.2$ for all simulations. As can be estimated from figures \ref{gtiles} and \ref{gwstiles}, the initial perturbation is located close to \scrip, so it is reached by the perturbation much earlier than the origin. 

\subsection{Implementation}


The implementation of the Einstein equations in the form given in \ssref{formulation}, the gauge conditions from \ssref{gauge} and the initial data described in \ssref{inidata} was performed in the NRPy+ code~\cite{Ruchlin:2017com}. NRPy+ is an open-source light-weight python infrastructure that creates optimized OpenMP-parallelized \texttt{C} code. It uses the method of lines with a 4th-order Runge-Kutta time integrator and 2nd order accurate finite differences were used for the spatial derivatives. Kreiss-Oliger dissipation, with strength $\epsilon_{\rm diss}=0.05$, is added for stability. 

An advantage of NRPy+ is that it allows to use Cartesian as well as different curvilinear coordinates. The latter is especially useful for hyperboloidal simulations, as time cuts of null infinity correspond to spheres, which are the natural boundary of spherical polar coordinates. This allows to use scri-fixing straightforwardly, as well as very efficiently change from spherical to axial symmetry and to 3D. To avoid evaluating where the coordinate system becomes singular (the origin and the axis), the code uses a staggered grid in the three spatial coordinates,~$(r,\theta,\phi)$. This means that the domain in for instance~$r \in (0,1)$ is divided in~$N_r$ cells, with a gridpoint in the center of each cell. The same is done for the~$\theta$ and~$\phi$ domains. This is standard in the~NRPy+ infrastructure. This also means that there are no gridpoints right on \scrip, where the formulation used is formally singular. The advantage is that no modifications need to be performed on the equations (see~\cite{\papgauge} for how to include a gridpoint on \scrip in spherical symmetry), while obtaining the signals at future null infinity requires to extrapolate half a radial step in post-processing. 

\begin{figure}
   \centering
    \includegraphics[width=0.99\linewidth]{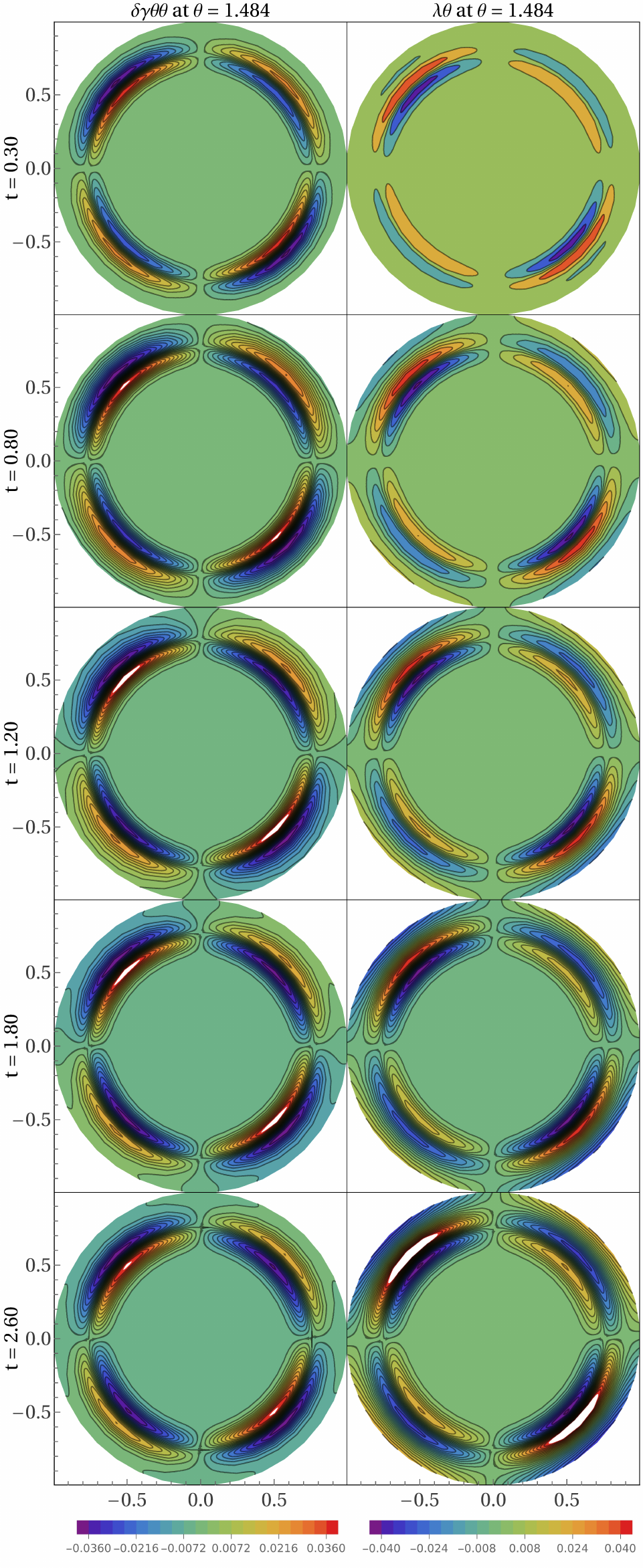}
    \caption{Snapshots of a 3D simulation with constraint-satisfying gauge wave initial data and evolved with gauge \eref{pbg}, showing variables $\delta\gamma_{\theta\theta}$ (see \eref{gththdef}) and $\lambda^\theta$ (defined at \eref{lthdef}) at polar angle $\theta=1.484$ for different times of the simulation. The outer circumference corresponds to \scrip. White parts take values beyond the range covered by the legend.}\label{gtiles}
\end{figure}
\begin{figure}
    \centering
    \includegraphics[width=0.99\linewidth]{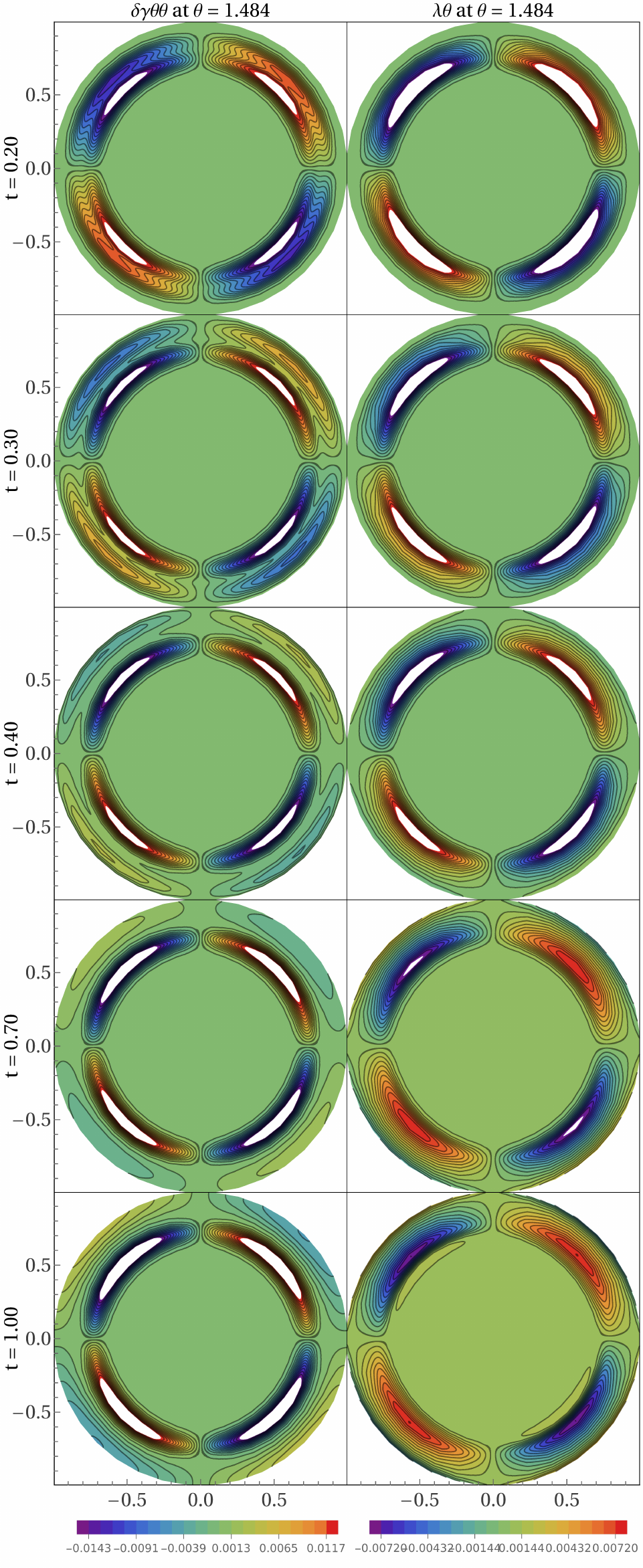}
    \caption{Snapshots of a 3D simulation with constraint-violating metric-perturbation initial data and evolved with gauge \eref{pbg}, showing variables $\delta\gamma_{\theta\theta}$ and $\lambda^\theta$ at polar angle $\theta=1.484$ for simulation times different to those in \fref{gtiles}. The outer circumference corresponds to \scrip. White parts take values beyond the range covered by the legend.}\label{gwstiles} 
\end{figure}


Despite the staggering of the grid and careful tensor rescalings (see~\cite{Ruchlin:2017com} for details), the code can blow up at the origin/$z$-axis even when a small time step is chosen in the evolutions. This has led to the choice of initial data with perturbations far away from the origin and close to \scrip (see figures \ref{gtiles} and \ref{gwstiles}), so evaluating the behaviour at future null infinity was possible before any blow-up took place in the central area. This is also the reason why no simulations including a black hole in puncture form at the center were attempted in this setup. Note that these limitations are code-related -- with a suitable infrastructure the formulation used can evolve spacetimes including black holes without time limitations. The results presented here are but a proof-of-concept that the evolution at \scrip works. 

\section{Results}\label{results}

The output of full 3D simulations is shown in \ssref{3d}, for both gauge wave and constraint-violating GW initial perturbations. As these evolutions are computationally expensive, convergence tests to check the reliability of the implementation are carried out in axial symmetry and presented in \ssref{conv}. 

The presentation of the results mostly show the quantity $\delta\gamma_{\theta\theta}$, the rescaled variation of the $\theta\theta$ component of the conformal spatial metric. When evolving in NRPy+, the latter is decomposed as 
\begin{equation}\label{gththdef}
\gamma_{\theta\theta} = \hat \gamma_{\theta\theta} + r^2\delta\gamma_{\theta\theta}
\end{equation}
when using spherical polar coordinates, to better treat coordinate singularities. 
The reason for focusing on this variable is that it provides information about the geometry of the angular sections of the spatial hypersurfaces, and as such it is closely related to GW radiation. 
This can be sketched as follows~\cite{bondisachsformalism}: assuming a Bondi frame asymptotically and $\Omega\sim1/\tilde r$ (as satisfied by~\eref{eq:compacfactor}), deviations of the angular metric from the unit-sphere geometry encode the trace-free Bondi shear through \scrip ($C_{AB}$)
\begin{equation}
\bar g_{AB}=\Omega^2\tilde g_{AB} \sim q_{AB}+\Omega C_{AB}+O(\Omega^2), 
\end{equation}
The derivative of the Bondi shear along the retarded time determines the GW news. Thus, $\delta\gamma_{\theta\theta}$ as a representative of the angular metric at \scrip is relevant for testing the code's treatment of gravitational radiation. 


\subsection{Simulations in full 3D}\label{3d}

The resolution for these runs is $N_r=360$, $N_\theta=18$ and $N_\varphi=36$, with enough gridpoints to capture all features smoothly. Figure~\ref{gtiles} shows $\delta\gamma_{\theta\theta}$ and $\lambda^\theta$, the latter defined from the BSSN/Z4 quantity $\Lambda^i$ in~\eref{Lambdadef} as 
\begin{equation}\label{lthdef}
\lambda^\theta = r\,\Lambda^\theta,
\end{equation}
for a simulation with an initial perturbation in the lapse that sends gauge waves across the spacetime. The contour plots show the variables evaluated at polar angle $\theta=1.484$ at different times. The radial coordinate extends from the origin at the center to $r=1$ where \scrip is located, while the revolution angle $\varphi$ covers $2\pi$. Part of the constraint-satisfying perturbation propagates outwards through \scrip and leaves the computational domain. 

\begin{figure} 
    \centering 
    \includegraphics[width=0.88\linewidth]{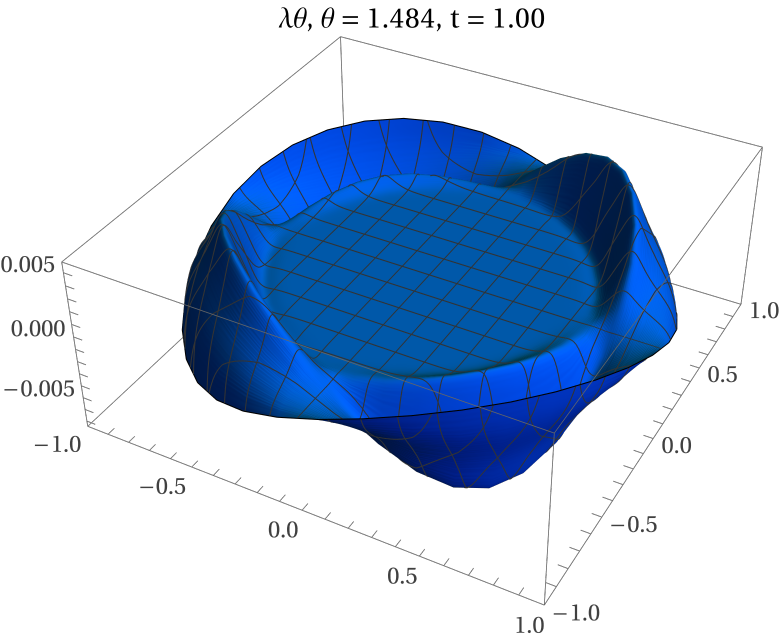}
    \includegraphics[width=0.88\linewidth]{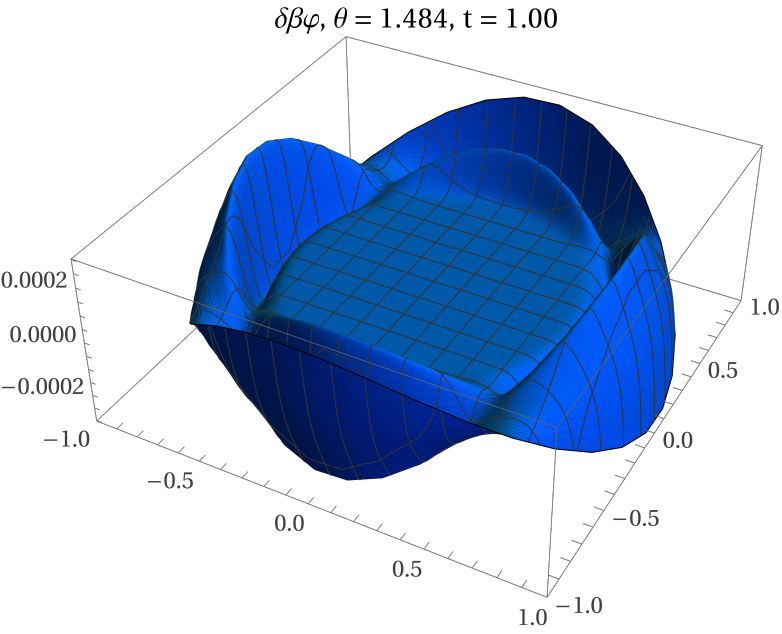}
    \caption{Snapshots at $t=1$ and for $\theta=1.484$ of 3D simulation with constraint-violating Brill-like GW initial data for $\lambda^\theta$ at the top and the $\varphi$ component of the shift $\beta^\varphi$ at the bottom. The compactified radial coordinate extends from the center to the outer circumference, where \scrip is located.}\label{3dplots}
\end{figure}

Closer to the goal of GW extraction are the results shown in \fref{gwstiles}, also showing $\delta\gamma_{\theta\theta}$ and $\lambda^\theta$, for constraint-violating Brill-like GW initial data. The perturbation, set in the diagonal components of the metric as described in~\ssref{inidata} excites all evolution quantities. The outwards-directed part of the GW radiation travels through \scrip.  
Figure~\ref{3dplots} shows 3D plots at $t=1$ and $\theta=1.484$ of $\lambda^\theta$, to complement the information provided by \fref{gwstiles}, and the shift component $\beta^\varphi$, to show the effect of the initial perturbation in a different variable. In both cases part of the GW content has already left through \scrip. 

\subsection{Convergence}\label{conv}

Given that NRPy+ is only OpenMP-parallelized, carrying out convergence tests in full 3D becomes very expensive and time-consuming. For this reason, convergence has been checked in axially symmetric simulations, using a 1.1 increase in resolution between runs. The number of gridpoints used in each coordinate direction are 
\vspace{-1ex}
\begin{itemize}\itemsep0ex
\item Low resolution: $N_r=450$ and $N_\theta=28$
\item Medium (Med) resolution: $N_r=494$ and $N_\theta=30$ 
\item High resolution: $N_r=544$ and $N_\theta=34$ 
\item Higher resolution: $N_r=600$ and $N_\theta=38$ 
\end{itemize}

\begin{figure}[h]
    \centering
    \includegraphics[width=0.99\linewidth]{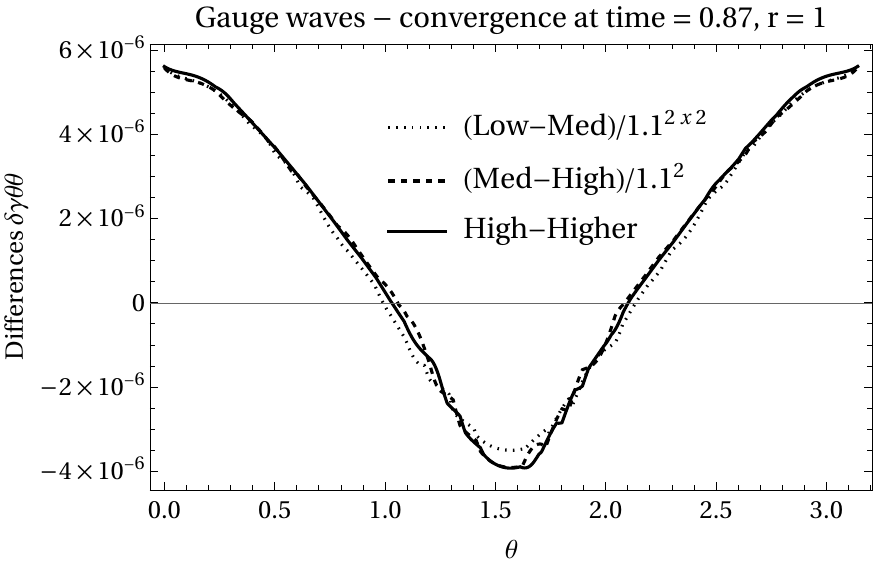}
    \includegraphics[width=0.99\linewidth]{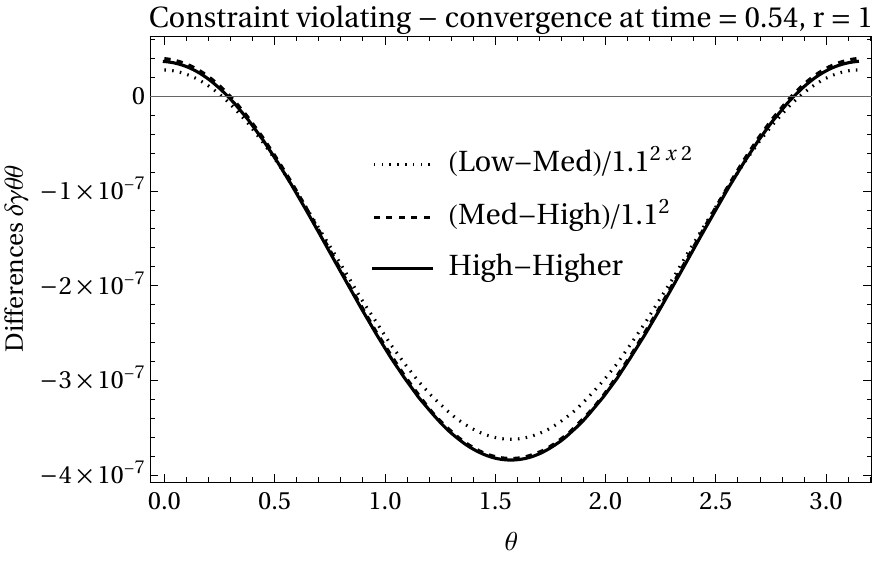}
    \caption{Pointwise convergence of $\delta\gamma_{\theta\theta}$ as a function of $\theta$ at \scrip: at $t=0.87$ for the gauge wave simulation at the top, and at $t=0.54$ for the constraint-violating Brill-like waves at the bottom. The differences are rescaled so that converging data are meant to lie on each other.}\label{pointwise}
\end{figure}

Pointwise convergence at \scrip for evolution variable $\delta\gamma_{\theta\theta}$ as a function of the polar angle $\theta$ given at a specific time is shown in \fref{pointwise}. The top figure corresponds to the runs with initial perturbation in the gauge, while the bottom one shows results from the constraint-violating runs with Brill-like GW initial data. The differences in $\delta\gamma_{\theta\theta}$ between runs are rescaled by the resolution increase to a power including the expected convergence order of 2. Coincidence between lines is good, meaning there is convergence. 

\begin{figure}[h!!!!] 
    \centering
    \includegraphics[width=0.99\linewidth]{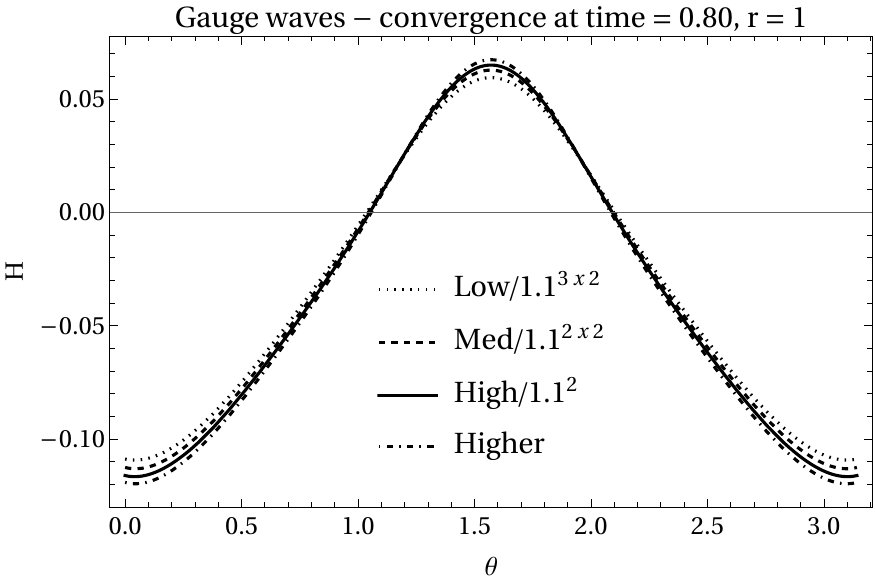}
    \includegraphics[width=0.99\linewidth]{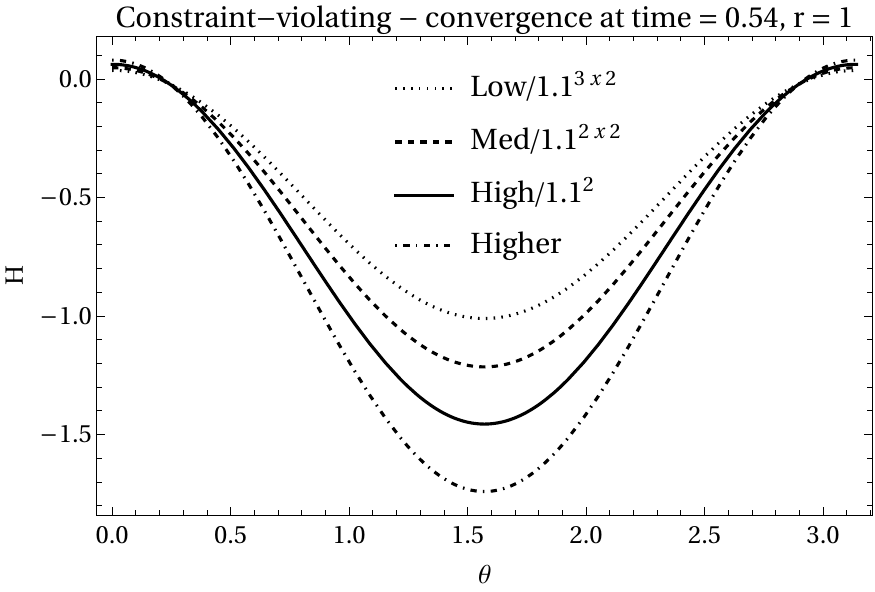}
    \caption{Pointwise convergence of the Hamiltonian constraint $H$ (given by (2.68a) in \cite{\alexthesis}) as a function of $\theta$ at \scrip: at $t=0.80$ for the gauge wave simulation at the top, and at $t=0.54$ for the constraint-violating Brill-like waves at the bottom. The values of $H$ are rescaled so that converging data are meant to lie on each other.}\label{Hconstr}
\end{figure}

Convergence of the constraints is a very good check for overall consistence of the implementation. Figure \ref{Hconstr} shows the values of the Hamiltonian constraint $H$ rescaled according to 2nd order convergence. These pointwise convergence plots are given as a function of $\theta$ for a specific time of the simulations. Convergence is good in the top plot presenting data from the constraint-satisfying gauge waves at the given time. However, the bottom plot corresponds to the runs with constraint-violating Brill-like GW initial perturbation, for which the constraints are not satisfied either initially or during the evolution. This is correctly recovered by the data, as the lines in the bottom plot clearly do not superpose. 

\begin{figure}[h!!!!] 
    \centering
    \includegraphics[width=0.99\linewidth]{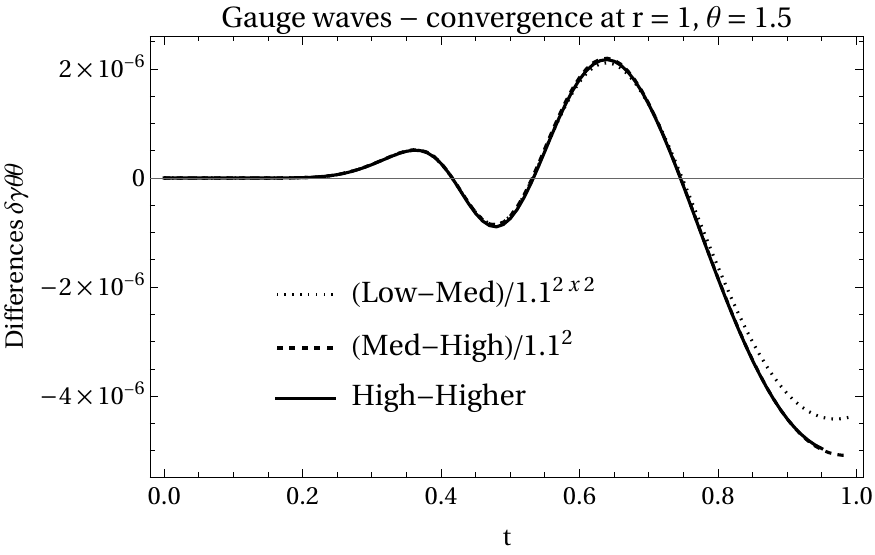}
    \includegraphics[width=0.99\linewidth]{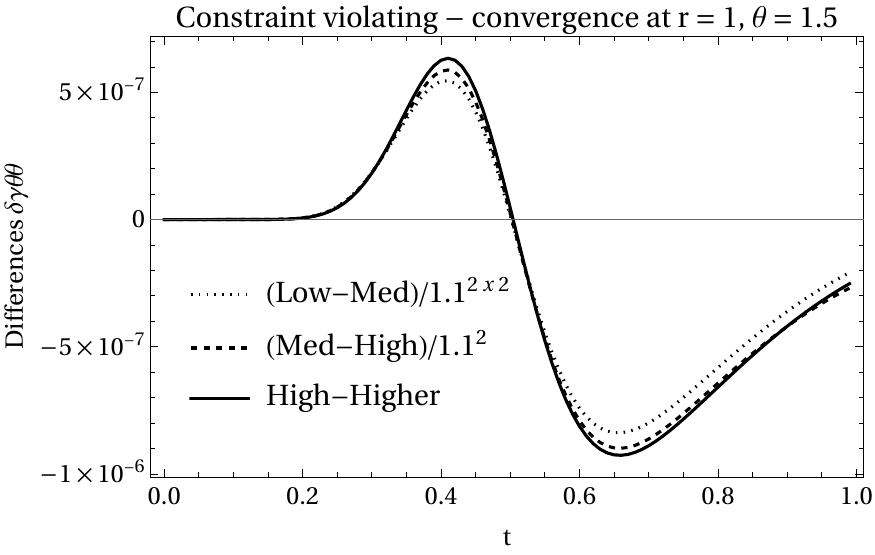}
    \caption{Convergence at $r=1$ (\scrip) and $\theta=1.5$ as a function of time. Plotted are the rescaled differences between $\delta\gamma_{\theta\theta}$ corresponding to the resolutions indicated in the legend, for gauge waves (top) and constraint-violating Brill-like GWs (bottom). Superposed lines indicate good convergence.}\label{pointwiseovert}
\end{figure}

The behaviour of the code's convergence over time is presented in the remaining figures. For the specific coordinate values of $r=1$ (\scrip) and $\theta=1.5$, \fref{pointwiseovert} shows the rescaled differences for $\delta\gamma_{\theta\theta}$ as a function of time, at the top for gauge waves and at the bottom for constraint-violating GWs. Convergence is quite good most of the time, and it is expected that the dashed and the solid lines coincide better than with the dotted line (using the lowest, least accurate resolution). 

\begin{figure}[h!!!!] 
    \centering
    \includegraphics[width=0.94\linewidth]{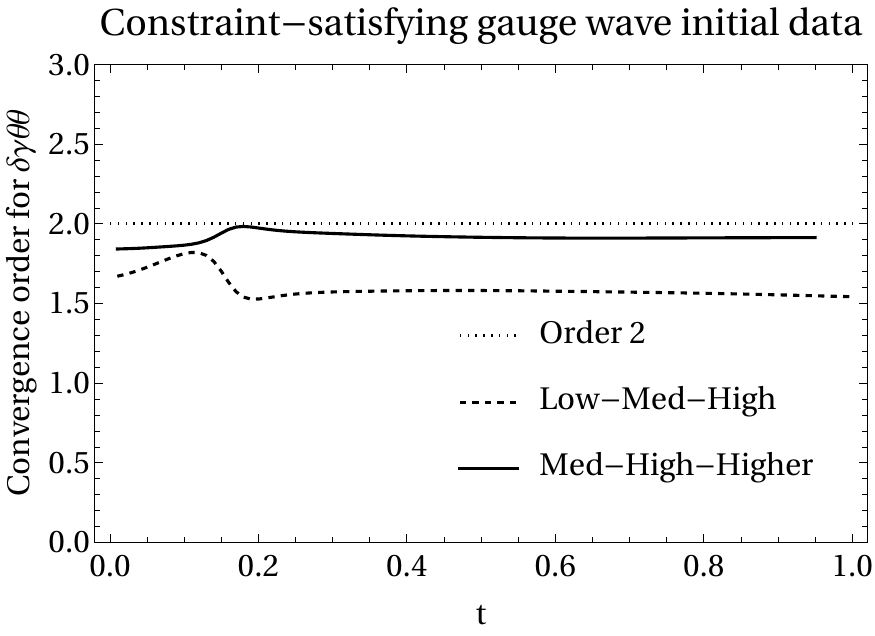}
    \includegraphics[width=0.94\linewidth]{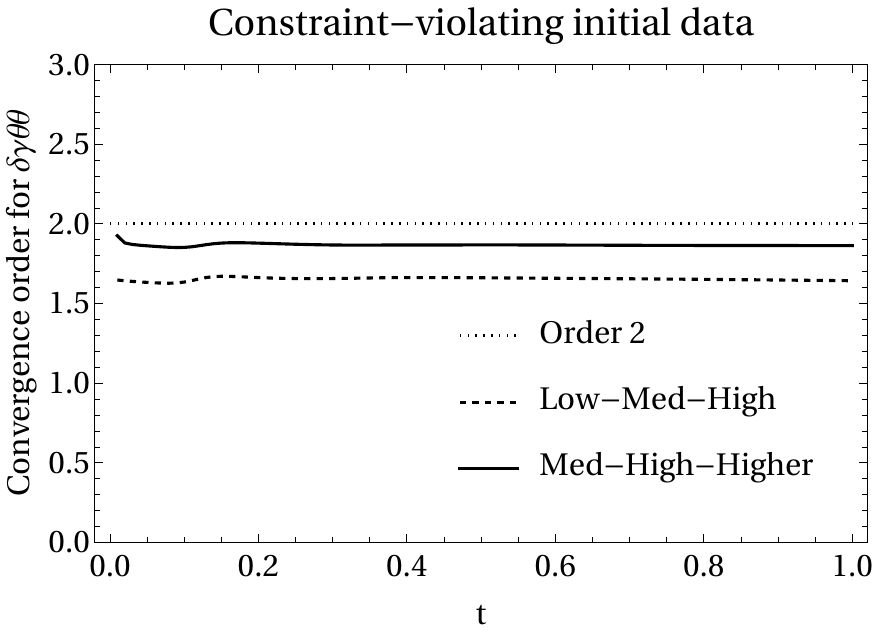}
    \caption{Convergence order for gauge waves (top) and constraint-violating GWs (bottom) determined from the norm of $\delta\gamma_{\theta\theta}$ over time, calculated as \eref{convorder}.}\label{norms}
\end{figure}

Finally, the L2-norm convergence of the simulations is presented in \fref{norms} for the $\delta\gamma_{\theta\theta}$ variable. The quantity shown corresponds to
\begin{equation}\label{convorder}
    \log_f\sqrt{(\sum (u_{f^2 h} - u_{fh})^2)/(\sum (u_{f h} - u_{h})^2)}\, ,
\end{equation}
where $u_{f^n h}$ corresponds to $\delta\gamma_{\theta\theta}$ in a simulation with a spatial step $f^n h$. The sum is performed on the whole grid. The $f=1.1$ logarithm of the quantity should then give the order of convergence of the code, which is 2.

The dashed line corresponds to the convergence order calculated via~\eref{convorder} using data from the low ($N_r=450$, $N_\theta=28$), medium ($N_r=494$, $N_\theta=30$) and high ($N_r=544$, $N_\theta=34$) resolution runs, while the solid line uses the medium, high and higher ($N_r=600$, $N_\theta=38$) resolutions. As expected, the second curve is much closer to the prescribed value of 2, as its data are further into the convergence regime than the other set, for which convergence is not properly achieved. 

\section{Discussion}\label{conclusions}

This work has used a full 3D numerical implementation of the Einstein equations along hyperboloidal slices, and is meant as a proof-of-concept prototype that can deal with signals at \scrip successfully. 
The implementation has been tested with simple configurations such a gauge waves and constraint-violating GWs, but it is also capable of evolving initial data including black holes without significant changes to the formulation. To the author's best knowledge, these are the first full 3D free hyperboloidal evolutions that use a common formulation like BSSN/Z4. 

The present implementation has technical limitations, such as the likely blowup at the coordinate singularities of the spherical polar coordinates used, and no MPI parallelization. This calls for the need of a more powerful code infrastructure to further develop and test hyperboloidal evolution. The plan for the present BSSN/Z4 formulation is to implement them in the Einstein Toolkit~\cite{steven_r_brandt_2026_short}, an actively-developed open-source numerical relativity code with a wide range of users. 

Showing the feasibility of the hyperboloidal method in absence of symmetries with a commonly used formulation in numerical relativity, the present work, is only the first step. Next in line is the derivation of hyperboloidal initial data without symmetries, covering constraint-satisfying GWs on a regular spacetime, as well as generic perturbations of black hole spacetimes. The main goal of this research line is to efficiently simulate binary systems of compact objects with accurate GW extraction (including gravitational memory) right at \scrip, of which the binary black hole scenario is currently the most relevant one. 
Obtaining reliable GW signals at \scrip will also require understanding and treating gauge effects in the hyperboloidal evolutions, as well as adapting the insights gained by CCE on BMS (Bondi-Metzner-Sachs) frames at \scrip~\cite{Mitman:2024uss}. 

\acknowledgments 

The author thanks Edgar Gasperin and Christian Peterson B\'orquez for valuable comments on the manuscript. He is also grateful to them, David Hilditch, An\i{}l Zengino\u{g}lu, J\"org Frauendiener and Sascha Husa for very useful discussions. 

The author is supported by the Ministry of Education and Universities through the Vicenç Mut program for the incorporation of research personnel into the science, technology and innovation ecosystem of the Balearic Islands (postdoctoral contracts 2024), with reference number POSTDOC2024\_20, funded by the Ministry of Education and Universities and by the European Union (ESF+), within the framework of the ESF+ Balearic Islands Program for the period 2021-2027.
This work was also supported by the Universitat de les Illes Balears (UIB) with funds from the Programa de Foment de la Recerca i la Innovació de la UIB 2024-2026 (supported by the yearly plan of the Tourist Stay Tax ITS2023-086); the Spanish Agencia Estatal de Investigación grants PID2022-138626NB-I00, RED2024-153978-E, RED2024-153735-E, funded by MICIU/AEI/10.13039/501100011033 and the ERDF/EU; and the Comunitat Autònoma de les Illes Balears through an ``Ajut per a projectes de recerca científica i tecnològic'' (PRD2025\_[099/100]) and through the Conselleria d'Educació i Universitats with funds from the European Union - European Regional Development Fund (ERDF) (SINCO2022/18146 - Plataforma HiTech-IAC3-BIO).
The author also thanks the Fundac\~ao para a  Ci\^encia e Tecnologia (FCT), Portugal, for the financial support to the Center for Astrophysics and Gravitation (CENTRA/IST/ULisboa) through the Grant Project~No.~UIDB/00099/2020. Funding with DOI 10.54499/DL57/2016/CP1384/CT0090 is graciously acknowledged. 

Most of the algebraic derivations were performed using the Mathematica package xAct \cite{xAct}. 
The equations were implemented in the open-source code NRPy+ \cite{Ruchlin:2017com}, and the author thanks its developers for their assistance. 

The author thankfully acknowledges the computer resources, technical expertise and assistance provided by CENTRA/IST. Some computations were performed at the cluster ``Baltasar-Sete-Sóis'' and supported by the H2020 ERC Consolidator Grant ``Matter and strong field gravity: New frontiers in Einstein's theory'' grant agreement no. MaGRaTh-646597.
This work was also made possible by the use of the Research Compute Hub (RCH) facilities (DOI:10.18124/CANTERBURYNZ-UCRCH, RRID:SCR\_027870) at the University of Canterbury.
The author also aknowledges the computational resources from UIB-IAC3's nura machine.  

AI assistance was used in the formatting of the equations in appendix~\ref{appendix}. 


\appendix



\section{Explicit form of gauge equations}\label{appendix}

The following are the explicit expressions of gauge conditions \eref{pbg}. Hatted quantities denote time-independent background quantities, and the evolution equation of the lapse is to be substituted into that of the shift, as the first term in the latter indicates. The expressions for $\tilde F^t$ and $\tilde F^i$ are given above in~\eref{ffuncs}. 
\begin{widetext}
\begin{subequations}
\begin{align}
\partial_t\alpha={}&
-\frac{\alpha^3 \tilde F^t}{\Omega^2}
-\frac{\alpha^2\Delta\tilde{K}}{\Omega}
-\frac{\alpha^2 \Kc}{\Omega}
-\frac{2\alpha^2\tilde \Theta }{\Omega}
+\frac{\alpha(\partial_i\hat\alpha)\hat\beta^i}{\hat\alpha}
+\beta^i\partial_i\alpha
-\frac{2\alpha(\partial_i\hat\alpha)\beta^i}{\hat\alpha}
-\frac{\alpha\hat\beta^i\partial_i\Omega}{\Omega}
+\frac{\alpha\beta^i\partial_i\Omega}{\Omega}
\nonumber \\
&-\frac{\alpha \Kc\hat\beta^i\hat\beta^j\hat\gamma_{ij}}{3\hat\alpha\hat\chi\Omega}
+\frac{2\alpha \Kc\hat\beta^j\beta^i\hat\gamma_{ij}}{3\hat\alpha\hat\chi\Omega}
-\frac{\alpha \Kc\beta^i\beta^j\hat\gamma_{ij}}{3\hat\alpha\hat\chi\Omega}
+\frac{\alpha^3 \chi \Kc\gamma^{ij}\hat\gamma_{ij}}{3\hat\alpha\hat\chi\Omega}.
\end{align}
\begin{align}
\partial_t\beta^i={}&
\frac{\beta^i}{\alpha}(\partial_t\alpha)
-\frac{\alpha^2 \tilde F^i}{\Omega^2}
+\frac{\alpha\Delta\tilde{K}\,\beta^i}{\Omega}
+\frac{\alpha \Kc\beta^i}{\Omega}
+\frac{2\alpha\tilde\Theta \beta^i}{\Omega}
+\alpha^2 \chi\,\bar\Lambda^i
+\hat\beta^j\partial_j\hat\beta^i
-2\beta^j\partial_j\hat\beta^i
\nonumber\\
&+\beta^j\partial_j\beta^i
-\frac{(\partial_j\hat\alpha)\hat\beta^i\hat\beta^j}{\hat\alpha}
+\frac{2(\partial_j\hat\alpha)\hat\beta^i\beta^j}{\hat\alpha}
-\frac{(\partial_j\alpha)\beta^i\beta^j}{\alpha}-\frac{\hat\beta^i\hat\beta^j\partial_j\hat\chi}{\hat\chi}
+\frac{\hat\beta^j\beta^i\partial_j\hat\chi}{\hat\chi}
+\frac{\hat\beta^i\beta^j\partial_j\hat\chi}{\hat\chi}
-\frac{\beta^i\beta^j\partial_j\hat\chi}{\hat\chi}
\nonumber \\
&-\frac{\hat\beta^i\hat\beta^j\partial_j\Omega}{\Omega}
+\frac{2\hat\beta^j\beta^i\partial_j\Omega}{\Omega}
-\frac{\beta^i\beta^j\partial_j\Omega}{\Omega}-\alpha \chi(\partial_j\alpha)\gamma^{ij}
+\hat\alpha\hat\chi(\partial_j\hat\alpha)\hat\gamma^{ij}
+\frac{\alpha^2}{2}(\partial_j \chi)\gamma^{ij}
\nonumber \\
&+\frac{\alpha^2 \chi}{\hat\chi}(\partial_j\hat\chi)\gamma^{ij}
+\frac{4\alpha^2 \chi}{\Omega}(\partial_j\Omega)\gamma^{ij}
-\frac{\hat\alpha^2\hat\chi}{\Omega}(\partial_j\Omega)\hat\gamma^{ij}
+\frac{\Kc\hat\beta^i\hat\beta^j\hat\beta^k\hat\gamma_{jk}}{3\hat\alpha\hat\chi\Omega}
-\frac{2\Kc\hat\beta^i\hat\beta^k\beta^j\hat\gamma_{jk}}{3\hat\alpha\hat\chi\Omega}
\nonumber \\
&+\frac{\Kc\hat\beta^i\beta^j\beta^k\hat\gamma_{jk}}{3\hat\alpha\hat\chi\Omega}
-\frac{\alpha^2 \chi \Kc\hat\beta^i\gamma^{jk}\hat\gamma_{jk}}{3\hat\alpha\hat\chi\Omega}
-\frac{2\hat\alpha \Kc\hat\beta^j\hat\gamma_{jk}\hat\gamma^{ik}}{3\Omega}
+\frac{2\hat\alpha \Kc\beta^j\hat\gamma_{jk}\hat\gamma^{ik}}{3\Omega}
+\hat\beta^j\hat\beta^k\hat\Gamma^i{}_{jk}
\nonumber \\
&-2\hat\beta^k\beta^j\hat\Gamma^i{}_{jk}
+\beta^j\beta^k\hat\Gamma^i{}_{jk}
+\frac{\hat\beta^j\hat\beta^k(\partial_l\hat\chi)\hat\gamma^{il}\hat\gamma_{jk}}{2\hat\chi}
-\frac{\hat\beta^k\beta^j(\partial_l\hat\chi)\hat\gamma^{il}\hat\gamma_{jk}}{\hat\chi}
+\frac{\beta^j\beta^k(\partial_l\hat\chi)\hat\gamma^{il}\hat\gamma_{jk}}{2\hat\chi}
+\frac{\hat\beta^j\hat\beta^k(\partial_l\Omega)\hat\gamma^{il}\hat\gamma_{jk}}{\Omega}
\nonumber \\
&-\frac{2\hat\beta^k\beta^j(\partial_l\Omega)\hat\gamma^{il}\hat\gamma_{jk}}{\Omega}
+\frac{\beta^j\beta^k(\partial_l\Omega)\hat\gamma^{il}\hat\gamma_{jk}}{\Omega}
-\frac{\alpha^2 \chi(\partial_j\hat\chi)\gamma^{kl}\hat\gamma_{kl}\hat\gamma^{ij}}{2\hat\chi}
-\frac{\alpha^2 \chi(\partial_j\Omega)\gamma^{kl}\hat\gamma_{kl}\hat\gamma^{ij}}{\Omega}.
\end{align}
\end{subequations}

\end{widetext}

\section*{References}
\bibliographystyle{unsrt}
\bibliography{../hypcomp}

\end{document}